\documentclass[conference]{IEEEtran}
\IEEEoverridecommandlockouts

\usepackage{amsmath,amssymb}
\usepackage{booktabs}
\usepackage{graphicx}
\usepackage{float}
\usepackage{capt-of}
\usepackage[caption=false,font=footnotesize]{subfig}
\usepackage{cite}
\usepackage{url}
\makeatletter
\renewcommand{\section}{\@startsection{section}{1}{\z@}%
  {1.0ex plus 0.4ex minus 0.2ex}%
  {0.45ex plus 0.2ex minus 0ex}%
  {\normalfont\normalsize\centering\scshape}}
\renewcommand{\subsection}{\@startsection{subsection}{2}{\z@}%
  {0.9ex plus 0.4ex minus 0.2ex}%
  {0.35ex plus 0.2ex minus 0ex}%
  {\normalfont\normalsize\itshape}}
\makeatother

\newcommand{\CN}{\mathcal{CN}}
\newcommand{\tr}{\operatorname{tr}}
\newcommand{\Grel}{G_{\mathrm{rel}}}
\newcommand{\CC}{\mathbf C_c}

\graphicspath{{./}}

\title{UAV Fluid-Antenna Channel Acquisition under Intra-Scan Channel Aging}

\author{
  \IEEEauthorblockN{
    Yuanhui Wu\textsuperscript{1},
    Hao Jiang\textsuperscript{2},
    Liang Wu\textsuperscript{3,4},
    Zaichen Zhang\textsuperscript{3,4}
  }
  \IEEEauthorblockA{
    \textsuperscript{1}College of Artificial Intelligence, Nanjing University of Information Science and Technology, China\\
    \textsuperscript{2}School of Cyber Science and Engineering, Southeast University, China\\
    \textsuperscript{3}National Mobile Communications Research Laboratory, Southeast University, China\\
    \textsuperscript{4}Purple Mountain Laboratories, Nanjing, China\\[-1mm]
    Emails: 202412621447@nuist.edu.cn, jiang.hao@seu.edu.cn, wuliang@seu.edu.cn, zczhang@seu.edu.cn\\
    Corresponding author: Hao Jiang\\[-1mm]
  }
}

\begin{document}
\IEEEaftertitletext{\vspace{-1.1em}}
\maketitle

\begin{abstract}
	Sequential sounding in UAV fluid-antenna systems (FASs) provides additional spatial information but delays transmission, causing earlier channel observations to age. This paper addresses the resulting information--freshness tradeoff by jointly determining where to probe and when to stop under blockwise hardware constraints. A dynamic Karhunen--Lo\`eve estimator aligns asynchronous measurements with the transmission state, covariance information gain selects feasible probe locations, and an age-discount identity with a local one-more-slot condition characterizes the covariance-level balance between information and freshness. Paired simulations show that temporal alignment recovers most of the lower-tail reliability lost by static stacking, covariance-aware probing ranks highest numerically among the evaluated policies, and optimized one-slot sounding becomes statistically competitive with two-slot alternatives at the highest tested mobility. Within the evaluated schedule family, increasing mobility shifts the competitive operating region toward shorter scans, supporting joint probe-placement and sounding-duration design.
\end{abstract}

\section{Introduction}
Low-altitude UAV motion causes the channel to evolve during sounding, so early measurements may be stale by the time data are transmitted~\cite{Yan2019UAVSurvey}. A fluid antenna system (FAS) exposes many candidate ports but probes only a few through limited radio-frequency (RF) hardware~\cite{Wong2023Preliminaries,Jiang2025UAVFAS}. Under block-parallel sounding, an additional slot collects one new measurement per hardware block, but it also postpones transmission and ages all previous observations. The central acquisition problem is therefore not simply to collect more ports, but to balance spatial information against observation freshness.

Existing FAS acquisition methods recover spatial information from partial observations through port selection, oversampling, KL inference, learned reconstruction, and spatial priors~\cite{Chai2022PortSelection,New2025Oversampling,Wu2026KLLAMP,Zhang2025GeoAngular,Zhang2026BeyondCovariance}. Other studies address outdated channel-state information (CSI), temporal learning, finite apertures, or blockwise switching~\cite{Psomas2023Outdated,Zou2024Online,Bi2026ActiveLearning,Zhang2026FiniteAperture,Wu2026Blockwise}. The closest timing-aware work uses a fixed set of measured ports to interpolate channels under switching delays~\cite{Dinis2026SpatioTemporal}, while statistical partial-FAS selection exploits incomplete historical CSI across ports and time~\cite{Han2026SemiBlind}. Neither addresses the joint decision considered here: choosing one feasible port per hardware block and deciding whether another sounding slot is worth its transmission delay.

This paper jointly determines where to probe and when to stop to improve lower-tail selected-port reliability at transmission. Dynamic-KL aligns asynchronous measurements with the transmission state, covariance IG selects feasible unused ports, and an age-discount identity with a local one-more-slot condition characterizes the information--freshness balance. Independent validation freezes task-level sounding budgets before testing.

Paired simulations isolate timing, geometry, and duration. Temporal alignment recovers most of the lower-tail loss from static stacking, covariance IG ranks highest numerically among the evaluated probing rules, and higher mobility favors shorter scans. At the highest tested mobility, optimized one-slot sounding is statistically competitive with the evaluated two-slot alternatives, supporting joint probe-placement and sounding-duration design without establishing a universally monotone budget rule.

The main contributions are as follows.
\begin{enumerate}
	\setlength{\itemsep}{0pt}
	\setlength{\parskip}{0pt}
	\setlength{\parsep}{0pt}
	\item We formulate block-parallel UAV-FAS acquisition as a joint probe-placement and sounding-duration problem with intra-scan channel evolution and transmission-time reliability.
	\item We combine transmission-state-aligned KL inference with covariance-IG probing and derive local covariance conditions that compare the information from another slot with its freshness penalty.
	\item Paired evaluations separate temporal alignment, probe geometry, and sounding duration, revealing the mobility-dependent region in which shorter scans become competitive.
\end{enumerate}

Sections~\ref{subsec:block-fas}--\ref{subsec:reliability} formulate the block-connected UAV--FAS model, mobility evolution, and transmission-time reliability objective. Sections~\ref{subsec:alignment}--\ref{subsec:one-more-slot} develop temporal state alignment, covariance-aware probing, and the one-more-slot condition. Section~\ref{sec:results} presents the numerical evaluation, and Section~\ref{sec:conclusion} concludes the paper.

{\em notation:} Boldface denotes matrices and vectors, while calligraphic letters denote sets. Superscripts $(\cdot)^T$ and $(\cdot)^H$ denote transpose and Hermitian transpose, $\tr(\cdot)$ is the trace, and $\CN(\boldsymbol\mu,\mathbf C)$ is a proper complex Gaussian distribution.

\begin{figure*}[!t]
	\centering
	\includegraphics[width=\textwidth]{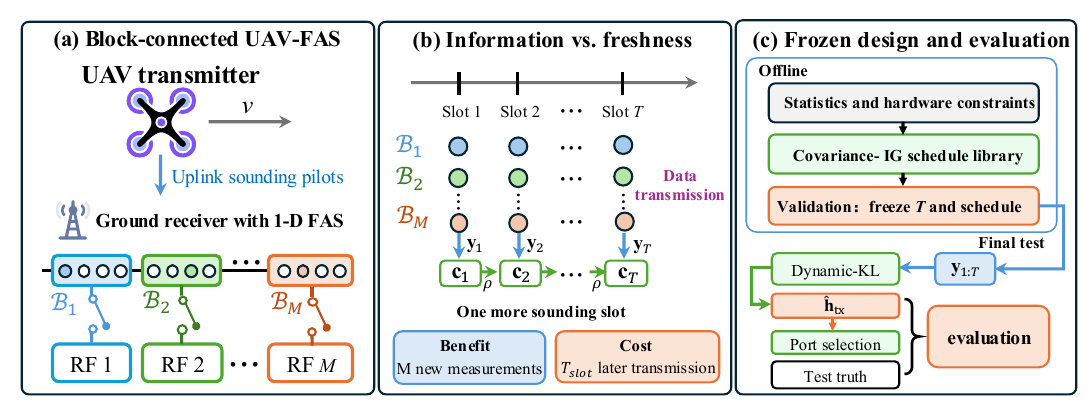}
	\caption{System overview and correspondence to the paper organization. (a) The block-connected uplink UAV--FAS model in Section~\ref{subsec:block-fas}. (b) The one-more-slot information--freshness tradeoff in Section~\ref{subsec:one-more-slot}: $M$ new measurements are obtained at the cost of delaying data transmission by $T_{\rm slot}$. (c) The frozen-design and final-test pipeline in Sections~\ref{subsec:reliability}, \ref{subsec:alignment}, \ref{subsec:probing}, and~\ref{sec:results}: covariance-IG schedules are generated offline, validation freezes $T$ and the schedule, and Dynamic-KL estimates the transmission-time channel before port selection; test truth is used only for evaluation.}
	\label{fig:overview}
\end{figure*}

\section{System Model and Problem Formulation}
\subsection{Block-connected FAS and spatial channel}
\label{subsec:block-fas}
Fig.~\ref{fig:overview} summarizes the block-parallel acquisition setting considered below.

Consider an uplink in which a UAV sends pilots to a ground receiver equipped with a one-dimensional FAS of aperture $L$. The $N$ candidate positions, $x_n=L(n-1)/(N-1)$ for $n=1,\ldots,N$, are partitioned into $M$ disjoint contiguous blocks $\{\mathcal B_b\}_{b=1}^{M}$. Each RF chain connects to one block and sounds one port per acquisition slot. If $\mathcal S_t$ is the slot-$t$ selection, then
\begin{equation}
 |\mathcal S_t\cap\mathcal B_b|=1,~
 \mathcal S_t\cap\mathcal S_\tau=\varnothing\ \ (t\ne\tau),
 \label{eq:hardware}
\end{equation}
where a port cannot be remeasured within an episode. A $T$-slot scan measures $m=MT$ ports.

The spatial model uses a directional Gaussian angle-of-arrival approximation with mean $\theta_0$ and standard deviation $\sigma_\theta$~\cite{Wu2026KLLAMP}. Letting $d_{ij}=(x_i-x_j)/\lambda$ and expressing $\sigma_\theta$ in radians, the covariance is
\begin{equation}
 [\mathbf R_h]_{ij}=e^{j2\pi d_{ij}\sin\theta_0}
 e^{-\frac12(2\pi|d_{ij}|\cos\theta_0\sigma_\theta)^2}.
 \label{eq:spatial}
\end{equation}
For $\mathbf R_h=\mathbf U_{\rm full}\boldsymbol\Lambda\mathbf U_{\rm full}^H$, set $\mathbf U=\mathbf U_{\rm full}(:,1\!:\!D)$. The rank-$D$ projected channel is
\begin{equation}
 \mathbf h_t=\mathbf U\mathbf c_t,~
 \mathbf c_t=\mathbf U^H\mathbf h_t,~\mathbf U\in\mathbb C^{N\times D}.
 \label{eq:kl}
\end{equation}
The same rank-$D$ projected channel is used by every estimator and reliability metric.

\subsection{Mobility-induced temporal evolution}
The coefficient trajectory follows
\begin{equation}
 \mathbf c_{t+1}=\boldsymbol\mu+\rho(\mathbf c_t-\boldsymbol\mu)
 +\sqrt{1-|\rho|^2}\,\mathbf q_t.
 \label{eq:ar1}
\end{equation}
where $\mathbf q_t\sim\CN(\mathbf0,\CC)$ is independent across slots and $\CC$ is the stationary coefficient covariance. The estimator adopts a zero-mean prior. For maximum Doppler $f_D=vf_c/c_0$, with UAV--receiver relative speed $v$, the one-slot diffuse correlation follows the Clarke/Jakes model,
\begin{equation}
 \rho=J_0(2\pi f_D T_{\rm slot}),~
 T_c=\frac{0.423}{f_D},~
 \eta=\frac{T_{\rm scan}}{T_c}=\frac{mT_{\rm slot}}{MT_c}.
 \label{eq:physics}
\end{equation}
Equation~\eqref{eq:ar1} is a first-order stationary Gauss--Markov surrogate matched at one slot. Its $L$-slot correlation is $\rho^L$, not the exact multi-lag value $J_0(2\pi f_DLT_{\rm slot})$. Scan time includes sounding but excludes switching and processing latency.

For $|\mathcal S_t|=M$, let $\mathbf P_{\mathcal S_t}\in\{0,1\}^{M\times N}$ denote the port-selection matrix. The observation is
\begin{equation}
 \mathbf y_t=\mathbf P_{\mathcal S_t}\mathbf h_t+\mathbf n_t
 =\mathbf A_t\mathbf c_t+\mathbf n_t,~
 \mathbf A_t=\mathbf P_{\mathcal S_t}\mathbf U,
 \label{eq:measurement}
\end{equation}
where $\mathbf n_t\sim\CN(\mathbf0,\sigma_n^2\mathbf I_M)$. The noise variance is calibrated once from the training split using a fixed reference set $\mathcal S_{\rm ref}$:
\begin{equation}
 \begin{aligned}
 p_{\rm ref}^{\rm tr}&=\mathop{\rm median}_{e\in\mathcal D_{\rm tr}}
 \frac{1}{|\mathcal S_{\rm ref}|}\sum_{n\in\mathcal S_{\rm ref}}|h_n^{(e)}|^2,\\
 \sigma_n^2&=p_{\rm ref}^{\rm tr}10^{-\Gamma_{\rm meas}/10}.
 \end{aligned}
 \label{eq:noise}
\end{equation}
The same $\sigma_n^2$ applies to every test episode, velocity, and sounding slot, so test realizations do not determine their own noise levels.

\subsection{Reliability metric and problem formulation}
\label{subsec:reliability}
With zero post-scan delay, $\mathbf h_{\rm tx}=\mathbf h_T$. Define
\begin{equation}
 \widehat n=\arg\max_n|\widehat h_{{\rm tx},n}|^2,~
 \Grel=\frac{|h_{{\rm tx},\widehat n}|^2}{\max_n|h_{{\rm tx},n}|^2}.
 \label{eq:grel}
\end{equation}
For episode $e$, transmission-time reconstruction error is
\begin{equation}
 \mathrm{NMSE}_{{\rm tx},e}=
 \frac{\|\widehat{\mathbf h}_{{\rm tx},e}-\mathbf h_{{\rm tx},e}\|_2^2}
 {\|\mathbf h_{{\rm tx},e}\|_2^2}.
 \label{eq:nmse}
\end{equation}
We report $10\log_{10}(E^{-1}\sum_e\mathrm{NMSE}_{{\rm tx},e})$. The primary metric is the empirical fifth percentile $Q_{0.05}(\Grel)$ over $E$ paired episodes. For data-link SNR $\Gamma_{\rm data}$ and $\gamma=10^{\Gamma_{\rm data}/10}$, the supporting rate metric is
\begin{equation}
 R_{\rm sel}=\log_2(1+\gamma|h_{{\rm tx},\widehat n}|^2),~
 R_5=Q_{0.05}(R_{\rm sel})\ \text{bit/s/Hz}.
 \label{eq:r5}
\end{equation}
This rate excludes sounding, switching, and processing overhead; it is a selected-link spectral-efficiency diagnostic, not net payload throughput. The reliability objective is
\begin{equation}
 \max_{T\in\{1,\ldots,\lfloor N/M\rfloor\},\{\mathcal S_t\}} Q_{0.05}(\Grel)
 ~\text{s.t.}~\eqref{eq:hardware}\text{--}\eqref{eq:measurement}.
 \label{eq:problem}
\end{equation}
Equation~\eqref{eq:problem} allows causal hardware-feasible schedules based on the known model and past observations. The experiments consider a finite budget set and offline covariance-generated schedules rather than the unrestricted combinatorial problem.

\section{Information--Freshness-Aware Acquisition}
\subsection{Temporal state alignment}
\label{subsec:alignment}
The endpoint estimator updates each observation at its acquisition time. With $\widehat{\mathbf c}_0^-=\mathbf0$ and $\mathbf C_0^-=\CC$, the proper-complex Dynamic-KL recursion is
\begin{equation}
\begin{aligned}
 \widehat{\mathbf c}_t^-&=\rho\widehat{\mathbf c}_{t-1}^+,\\
 \mathbf C_t^-&=|\rho|^2\mathbf C_{t-1}^++(1-|\rho|^2)\CC,\\
 \mathbf V_t&=\mathbf A_t\mathbf C_t^-\mathbf A_t^H+\sigma_n^2\mathbf I,\\
 \mathbf K_t&=\mathbf C_t^-\mathbf A_t^H\mathbf V_t^{-1},\\
 \widehat{\mathbf c}_t^+&=\widehat{\mathbf c}_t^-+
 \mathbf K_t(\mathbf y_t-\mathbf A_t\widehat{\mathbf c}_t^-).
\end{aligned}
\label{eq:kalman}
\end{equation}
The Joseph covariance form
\begin{equation}
 \mathbf C_t^+=(\mathbf I-\mathbf K_t\mathbf A_t)\mathbf C_t^-
 (\mathbf I-\mathbf K_t\mathbf A_t)^H+\sigma_n^2\mathbf K_t\mathbf K_t^H
 \label{eq:joseph}
\end{equation}
preserves Hermitian positive semidefiniteness, and $\widehat{\mathbf h}_{\rm tx}=\mathbf U\widehat{\mathbf c}_T^+$. The compared estimators and their information sets are specified in Section~\ref{sec:results}.

\subsection{Covariance-aware feasible probing}
\label{subsec:probing}
For candidate set $\mathcal S$, let $\mathbf C_t^+(\mathcal S)$ denote the posterior covariance. The slot objective is the log-determinant reduction
\begin{equation}
 \Delta I_t(\mathcal S)=\log\det\mathbf C_t^--\log\det\mathbf C_t^+(\mathcal S),
 \label{eq:ig}
\end{equation}
a D-optimal uncertainty criterion. Define $\mathbf u_n=\mathbf U^H\mathbf e_n\in\mathbb C^D$. At each greedy step, the eligible port with the largest exact rank-one increment
\begin{equation}
 \Delta I_t(n)=\log\!\left(1+\frac{\mathbf u_n^H\mathbf C\mathbf u_n}{\bar\sigma_n^2}\right)
 \label{eq:rankone}
\end{equation}
is selected. Its block is then closed, used ports remain ineligible, and $\mathbf C$ receives a rank-one update. The design and Kalman correction use the same training-calibrated variance $\bar\sigma_n^2=\sigma_n^2$ from \eqref{eq:noise}. Since covariance evolution is value-independent, each $(\rho,T)$ schedule is computed once offline.

The probing comparison fixes the first slot, measurement count, estimator, and paired episodes. Fixed repeats the saved pattern, whereas Random samples uniformly. Covariance variance ranks $[\mathbf U\mathbf C_t^-\mathbf U^H]_{nn}$ without conditional updates, and Gain-only ranks $|\widehat h_{t,n}^-|^2$. Gain--uncertainty uses $s_{t,n}=|\widehat h_{t,n}^-|^2+\beta[\mathbf U\mathbf C_t^-\mathbf U^H]_{nn}^{1/2}$, with $\beta$ selected on validation data. Every policy selects one unused port per block without true-channel access. One-shot IG-8 designs one feasible slot by covariance IG and transmits immediately afterward.

For a finite admissible budget set $\mathcal B$, validation data form the pointwise bootstrap-screened competitive set $\mathcal C_{\rm val}(v)$. The operating point is
\begin{equation}
 m_{\rm op}(v)=\min\mathcal C_{\rm val}(v).
 \label{eq:budget}
\end{equation}
The choice is frozen before final testing. Validation and test seeds are disjoint, conditions are paired within each split, and disjoint validation blocks measure selection variability. Online costs are $O(TM)$ for schedule access, $O(D^2M+DM^2+M^3)$ per update, and $O(ND+N)$ for reconstruction and search; offline scheduling costs $O(TMND^2)$.

\subsection{One-more-slot uncertainty condition}
\label{subsec:one-more-slot}
Propagating $\mathbf C_t^+$ through $L$ unobserved transitions under \eqref{eq:ar1} gives
\begin{equation}
 \mathbf C_{t+L|t}=|\rho|^{2L}\mathbf C_t^++(1-|\rho|^{2L})\CC,
 \label{eq:discount}
\end{equation}
so $\CC-\mathbf C_{t+L|t}=|\rho|^{2L}(\CC-\mathbf C_t^+)$. The value of an earlier posterior therefore decays geometrically as transmission moves farther from its measurement time.

Let $\mathbf D_{t+1}=\mathbf C_{t+1}^--\mathbf C_{t+1}^+$ denote the covariance reduction from the next feasible slot. The recursion gives
\begin{equation}
 \mathbf C_{t+1}^+-\mathbf C_t^+=(1-|\rho|^2)(\CC-\mathbf C_t^+)-\mathbf D_{t+1},
 \label{eq:trace-change}
\end{equation}
and another slot reduces the endpoint posterior trace exactly when
\begin{equation}
 \tr(\mathbf D_{t+1})>(1-|\rho|^2)\tr(\CC-\mathbf C_t^+).
 \label{eq:benefit}
\end{equation}
The condition compares the covariance reduction from a new slot with the freshness penalty of delayed transmission. It is not a guarantee or predictor of task-level $Q_{0.05}(\Grel)$ improvement, which is evaluated numerically.

For a prescribed feasible schedule, define the first locally non-beneficial extension, when it exists, as
\begin{equation}
 \begin{aligned}
 T_{\rm lnb}&=\min\!\left\{t:\tr(\mathbf D_{t+1})\leq
 (1-|\rho|^2)\tr(\CC-\mathbf C_t^+)\right\},\\
 m_{\rm lnb}&=MT_{\rm lnb}.
 \end{aligned}
 \label{eq:cov-transition}
\end{equation}
Stronger aging increases the freshness term, whereas noise and redundant ports reduce $\mathbf D_{t+1}$. The index is local to the prescribed continuation and need not be globally optimal; nonmonotone feasible sets rule out a universal single-crossing threshold. As $|\rho|\to1$, the freshness term vanishes. As $|\rho|\to0$, prediction returns to $\CC$ and erases earlier uncertainty reductions.

% These double-column floats are declared before the results heading so that
% IEEEtran can place them at the tops of the two result pages rather than after
% the conclusion. Their captions and all discussion remain in Section IV.
\begin{figure*}[!t]
\centering
\vspace*{3pt}
\subfloat[\hspace{1.0em}Transmission-time NMSE.\label{fig:estimator-nmse}]{%
\includegraphics[width=.325\textwidth,trim=12pt 0 0 0,clip]{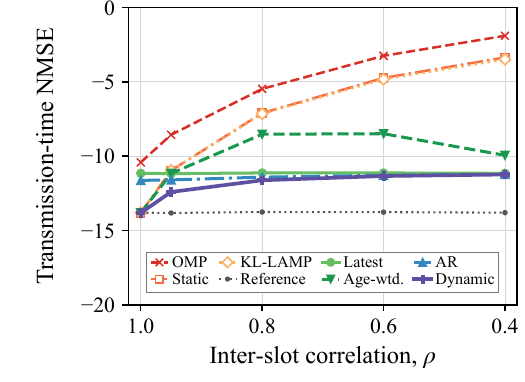}}
\hfill
\subfloat[\hspace{1.0em}Fifth-percentile retained gain.\label{fig:estimator-q05}]{%
\includegraphics[width=.325\textwidth,trim=12pt 0 0 0,clip]{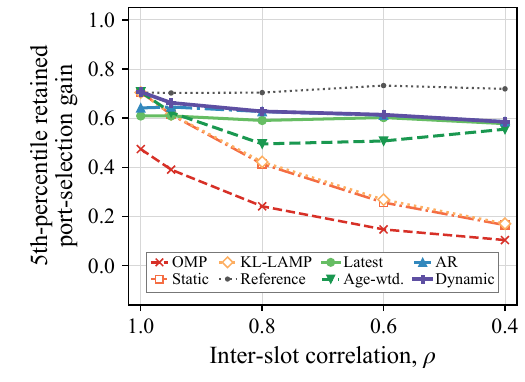}}
\hfill
\subfloat[\hspace{1.0em}Fifth-percentile achievable rate.\label{fig:estimator-r5}]{%
\includegraphics[width=.325\textwidth,trim=12pt 0 0 0,clip]{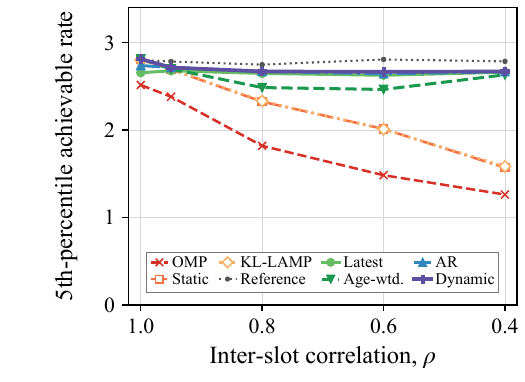}}
\caption{Estimator comparison under a fixed two-slot, 16-port schedule with $N=200$, $M=8$, $D=12$, 10-dB measurement/data SNR, and 4,000 paired test episodes per correlation point. (a) Transmission-time NMSE. (b) Fifth-percentile retained gain. (c) Fifth-percentile selected-link rate. Static snapshot estimators degrade as the inter-slot correlation decreases, whereas freshness-aware estimators remain close.}
\label{fig:estimator-comparison}
\vspace{-2mm}
\end{figure*}

\begin{figure}[!t]
\centering
\subfloat[Fifth-percentile retained gain.\label{fig:probing-q05}]{%
\includegraphics[width=.49\columnwidth,trim=12pt 0 0 0,clip]{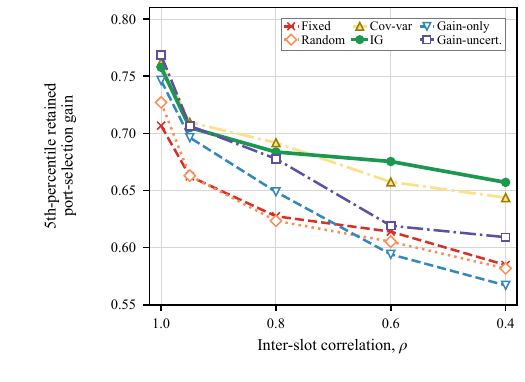}}
\hfill
\subfloat[Fifth-percentile achievable rate.\label{fig:probing-r5}]{%
\includegraphics[width=.49\columnwidth,trim=12pt 0 0 0,clip]{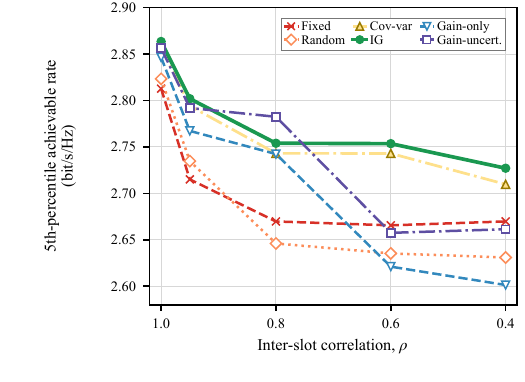}}
\caption{Post-anchor probing with $m=16$, $T=2$, a common first slot, Dynamic-KL inference, and 4,000 paired test episodes per correlation point. Each policy selects one unused port per block in slot two. (a) Fifth-percentile retained gain. (b) Fifth-percentile selected-link rate.}
\label{fig:probing-comparison}
\vspace{-2mm}
\end{figure}

\begin{figure*}[!t]
\centering
\subfloat[Useful sounding-depth landscape.\label{fig:budget-landscape}]{%
\includegraphics[width=.325\textwidth]{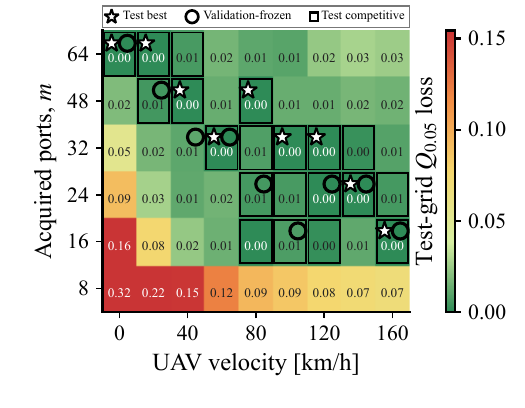}}
\hfill
\subfloat[Native-time retained-gain reliability.\label{fig:budget-q05}]{%
\includegraphics[width=.325\textwidth]{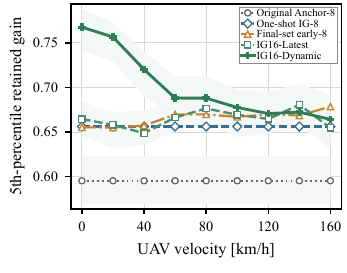}}
\hfill
\subfloat[Native-time rate reliability.\label{fig:budget-r5}]{%
\includegraphics[width=.325\textwidth]{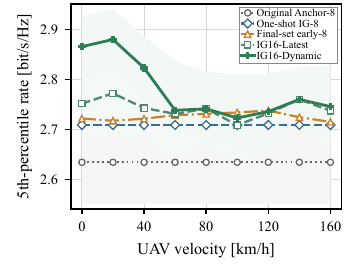}}
\caption{Mobility contracts the advantage of sequential sounding within the evaluated schedule family. (a) Test-grid loss $\Delta Q_{0.05}(m,v)=Q_{0.05}^{\rm best}(v)-Q_{0.05}(m,v)$ for $m\in\{8,16,24,32,48,64\}$ at 28 GHz and $T_{\rm slot}=66.67\,\mu$s, using 20,000 validation and 50,000 test episodes per cell. Stars, circles, and boxes denote test bests, validation-frozen operating points, and test-competitive budgets. (b)--(c) Native-completion reliability for five methods over 4,000 paired episodes per velocity; shaded regions are pointwise 95\% intervals.}
\label{fig:budget-decomposition}
\vspace{-2mm}
\end{figure*}

\section{Numerical Results}
\label{sec:results}
\paragraph{Simulation setup and protocol}
Table~\ref{tab:parameters} summarizes the fixed parameters and evaluated grids. The rank-12 KL basis retains 99.59\% of the covariance energy. Statistics, validation, and test sets are disjoint; validation selects all checkpoints and hyperparameters before paired final testing. Figure captions identify the controlled variable and native transmission time.

\begin{table}[!t]
\caption{Simulation parameters and evaluation protocol.}
\label{tab:parameters}
\centering
\setlength{\tabcolsep}{2.5pt}
\renewcommand{\arraystretch}{1.04}
\begin{tabular}{@{}p{0.29\columnwidth}p{0.67\columnwidth}@{}}
\toprule
Parameter & Value \\
\midrule
FAS geometry & $N=200$ ports over $4\lambda$; KL rank $D=12$ \\
RF hardware & $M=8$ chains/blocks; 25 ports per block \\
Channel statistics & $\theta_0=20^\circ$, $\sigma_\theta=10^\circ$; first-order Gauss--Markov \\
Correlation grids & Fig.~\ref{fig:estimator-comparison}: $\{1,.99,.95,.9,.8,.6,.4\}$; Fig.~\ref{fig:probing-comparison}: $\{1,.95,.8,.6,.4\}$ \\
Measurement/data SNR & $10$ dB / $10$ dB \\
Noise calibration & $|\mathcal S_{\rm ref}|=16$; $\sigma_n^2=0.0865248$ \\
Data split & 12,000 statistics / 4,000 validation / 4,000 test \\
KL-LAMP training & 10 layers; 200 epochs; batch 256; AdamW $(2\!\times\!10^{-3},10^{-4})$; 0--30-dB SNR \\
Selected settings & Angular-OMP sparsity 3; Gain--uncertainty $\beta=4$ \\
Mobility/budget grid & 28 GHz; $66.67\,\mu$s; $v=\{0,20,\ldots,160\}$ km/h; $m=\{8,16,24,32,48,64\}$ \\
Budget audit & 20,000 validation / 50,000 test per cell; five 4,000-sample validation blocks \\
Uncertainty & 500/1000-replicate pointwise paired bootstrap \\
\bottomrule
\end{tabular}
\end{table}

\paragraph{Temporal alignment under intra-scan aging}
Fig.~\ref{fig:estimator-comparison} isolates temporal inference by fixing the two-slot probing geometry and paired episodes. The static reference, Static KL-LMMSE, and frozen KL-LAMP~\cite{Wu2026KLLAMP} treat the acquired samples as a snapshot, whereas Latest-slot, Age-weighted, AR-aided, and Dynamic-KL use the same samples with progressively richer treatment of their measurement ages. Static stacking degrades as $\rho$ decreases, whereas latest-slot, AR-aided, and Dynamic-KL estimators remain stable by respecting measurement age. At $\rho=0.4$, KL-LAMP gives $-3.480$ dB NMSE and $Q_{0.05}(\Grel)=0.170$; the three freshness-aware estimators reach retained gains of 0.578, 0.586, and 0.585.

\paragraph{Covariance-aware probing}
Fig.~\ref{fig:probing-comparison} isolates post-anchor port selection: Dynamic-KL, the first-slot measurements, and paired episodes are fixed, while the second-slot rule is varied. Fixed and Random are geometry-agnostic controls; the remaining policies score variance, gain, both, and conditional IG, respectively. Validation selects $\beta=4$ for Gain--uncertainty from $\{0,0.25,0.5,1,2,4\}$. At $\rho=0.4$, the retained-gain fifth percentiles for Fixed, Random, covariance variance, IG, gain-only, and Gain--uncertainty are 0.5848, 0.5818, 0.6437, 0.6571, 0.5670, and 0.6091. IG also gives the highest reported rate, 2.727 bit/s/Hz, supporting covariance-aware geometric coverage.

\paragraph{Mobility and sounding duration}
Fig.~\ref{fig:budget-decomposition}\subref{fig:budget-landscape} evaluates six budgets from 0 to 160 km/h. The validation-frozen sequence is $m=\{64,48,32,32,24,16,24,24,16\}$ for $v=\{0,20,\ldots,160\}$ km/h. It is test-competitive at eight of nine velocities; the sole regret is 0.0109 at 40 km/h. Mobility therefore contracts the favorable depth region overall, without producing a strictly monotone sequence.

Figs.~\ref{fig:budget-decomposition}\subref{fig:budget-q05}--\subref{fig:budget-r5} and Table~\ref{tab:one-shot} separate sounding duration from transmission timing under native completion. Anchor-8 uses fixed geometry; One-shot IG-8 designs one feasible covariance-IG slot; FinalSet-early-8 advances the precomputed second IG16 geometry as a timing control; and the two IG16 methods share geometry but differ in temporal inference. One-slot methods transmit at $t=1$, two-slot methods at $t=2$, with common Jakes mapping, channels, noise, and SNR.

\begin{table}[!t]
\caption{Native-completion comparison at 160 km/h ($\rho=0.3755$), 28 GHz, and $T_{\rm slot}=66.67\,\mu$s.}
\label{tab:one-shot}
\centering
\scriptsize
\setlength{\tabcolsep}{2.3pt}
\begin{tabular*}{\columnwidth}{@{\extracolsep{\fill}}lrrr@{}}
\toprule
Method & NMSE [dB] & $Q_{0.05}(\Grel)$ & $R_5$ \\
\midrule
Original Anchor-8 & $-11.410$ & 0.5953 & 2.6347 \\
One-shot IG-8 & $-11.545$ & 0.6561 & 2.7092 \\
FinalSet-early-8 & $-11.602$ & \textbf{0.6785} & 2.7155 \\
IG16-Latest & $-11.640$ & 0.6546 & 2.7377 \\
IG16-Dynamic & $\mathbf{-11.692}$ & 0.6642 & \textbf{2.7461} \\
\bottomrule
\end{tabular*}
\end{table}

At 160 km/h, one-shot IG-8 differs from IG16-Latest and IG16-Dynamic by 0.0015 (95\% interval $[-0.0369,0.0290]$) and $-0.0081$ ($[-0.0449,0.0236]$), respectively, but improves on Original Anchor-8 by 0.0608 ($[0.0383,0.0839]$). FinalSet-early-8 reaches the largest tabulated $Q_{0.05}$, 0.6785, isolating the timing benefit of its prescribed geometry rather than an independently optimized one-shot policy. Dynamic-KL retained gain falls from 0.7676 to 0.6642; its advantage over one-shot IG-8 is resolved through 80 km/h for retained gain and 40 km/h for rate. Thus, informative one-slot geometry improves the anchor and becomes competitive with two-slot sounding at high mobility.

\section{Conclusion}
\label{sec:conclusion}
Intra-scan channel aging creates a core trade-off between probe placement and observation freshness in UAV fluid antenna systems. Covariance-IG scheduling optimizes time-varying channel probing, and channel aging shortens the cost-effective scan window. At 160 km/h, rapid aging offsets the gain of an extra slot, so the one-shot IG-8 scheduler achieves comparable performance to two-slot schemes. Future work will extend the framework with aperture coupling, multi-lag channel evolution, hardware latency and real UAV channel data.


\begin{thebibliography}{99}

\bibitem{Yan2019UAVSurvey}
C.~Yan, L.~Fu, J.~Zhang, and J.~Wang, ``A comprehensive survey on {UAV} communication channel modeling,'' \emph{IEEE Access}, vol.~7, pp. 107769--107792, 2019.

\bibitem{Wong2023Preliminaries}
K.-K. Wong, W.~K. New, H.~Xu, K.-F. Tong, and C.-B. Chae, ``Fluid antenna system---part {I}: Preliminaries,'' \emph{IEEE Communications Letters}, vol.~27, no.~8, pp. 1919--1923, August 2023.

\bibitem{Jiang2025UAVFAS}
H.~Jiang, W.~Shi, Z.~Chen, Z.~Zhang, K.-K. Wong, and H.~Shin, ``Dynamic channel modeling of fluid antenna systems in {UAV} communications,'' \emph{IEEE Wireless Communications Letters}, vol.~14, no.~10, pp. 3169--3173, October 2025.

\bibitem{Chai2022PortSelection}
Z.~Chai, K.-K. Wong, K.-F. Tong, Y.~Chen, and Y.~Zhang, ``Port selection for fluid antenna systems,'' \emph{IEEE Communications Letters}, vol.~26, no.~5, pp. 1180--1184, May 2022.

\bibitem{Zhang2025GeoCSI}
Z.~Zhang, D.~Morales-Jim{\'e}nez, J.~Dang, Z.~Zhang, C.~Masouros, and H.~Jiang, ``Low-complexity {CSI} acquisition exploiting geographical diversity in fluid antenna system,'' in \emph{IEEE Global Communications Conference Workshops}, 2025, pp. 308--313.

\bibitem{New2025Oversampling}
W.~K. New, K.-K. Wong, H.~Xu, F.~R. Ghadi, R.~D. Murch, and C.-B. Chae, ``Channel estimation and reconstruction in fluid antenna system: Oversampling is essential,'' \emph{IEEE Transactions on Wireless Communications}, vol.~24, no.~1, pp. 309--322, January 2025.

\bibitem{Wu2026KLLAMP}
Y.~Wu, Z.~Zhang, H.~Jiang, K.-K. Wong, and C.-B. Chae, ``Learned-approximate message passing under Karhunen--Lo\`eve modeling for fluid antenna systems,'' \emph{IEEE Wireless Communications Letters}, vol.~15, pp. 2719--2723, 2026.

\bibitem{Zhang2025GeoAngular}
Z.~Zhang, J.~Dang, D.~Morales-Jim{\'e}nez, H.~Jiang, Z.~Zhang, C.~Masouros, and C.-B. Chae, ``Joint activity detection and channel estimation for fluid antenna system exploiting geographical and angular information,'' arXiv:2512.15342, 2025.

\bibitem{Elganimi2026Critical}
T.~Y. Elganimi, P.~Ram{\'i}rez-Espinosa, D.~Morales-Jim{\'e}nez, and F.~J. L{\'o}pez-Mart{\'i}nez, ``Channel estimation and reconstruction in fluid antenna multiple access: Myths, misconceptions and critical questions,'' arXiv:2606.01842, 2026.

\bibitem{Liang2025NNReconstruction}
H.~Liang, Z.~Zhang, J.~Dang, H.~Jiang, and Z.~Zhang, ``Neural networks-enabled channel reconstruction for fluid antenna systems: A data-driven approach,'' arXiv:2511.14520, 2025.

\bibitem{Zhang2026BeyondCovariance}
Z.~Zhang, H.~Jiang, K.-K. Wong, H.~Shin, and R.~D. Murch, ``Beyond covariance: Generative spatial correlation modeling and channel interpolation for fluid antenna systems,'' arXiv:2604.16639, 2026.

\bibitem{Han2026SemiBlind}
T.~Han, Y.~Zhu, K.-K. Wong, G.~Zheng, C.-B. Chae, and X.~You, ``Semi-blind fluid antenna system: Port selection via statistical analysis,'' arXiv:2608.08339, 2026.

\bibitem{Zhang2026GeometryStructured}
Z.~Zhang, K.-K. Wong, K.~Meng, D.~Morales-Jim{\'e}nez, H.~Jiang, C.~Masouros, H.~Shin, and Z.~Zhang, ``Geometry-structured channel reconstruction for conventional and fluid antenna systems: Bayesian inference and fundamental limits,'' arXiv:2606.04001, 2026.

\bibitem{Zhang2023FastPort}
S.~Zhang, J.~Mao, Y.~Hou, Y.~Chen, K.-K. Wong, Q.~Cui, and X.~Tao, ``Fast port selection using temporal and spatial correlation for fluid antenna systems,'' in \emph{IEEE Statistical Signal Processing Workshop}, July 2023, pp. 95--99.

\bibitem{Psomas2023Outdated}
C.~Psomas, G.~M. Kraidy, K.-K. Wong, and I.~Krikidis, ``Fluid antenna systems with outdated channel estimates,'' in \emph{IEEE International Conference on Communications}, May 2023, pp. 2970--2975.

\bibitem{Zou2024Online}
J.~Zou, S.~Sun, and C.~Wang, ``Online learning-induced port selection for fluid antenna in dynamic channel environment,'' \emph{IEEE Wireless Communications Letters}, vol.~13, no.~2, pp. 313--317, February 2024.

\bibitem{Bi2026ActiveLearning}
Y.~Bi and D.~H.~K. Tsang, ``Online active learning for adaptive channel estimation in fluid antenna systems,'' in \emph{IEEE International Conference on Communications}, July 2026, pp. 1--6.

\bibitem{Zhang2026FiniteAperture}
Z.~Zhang, K.-K. Wong, H.~Jiang, F.~R. Ghadi, H.~Shin, and Y.~Zhang, ``Finite-aperture fluid antenna array design: Analysis and algorithm,'' \emph{IEEE Wireless Communications Letters}, vol.~15, pp. 3199--3203, 2026.

\bibitem{Wu2026Blockwise}
Y.~Wu, Z.~Zhang, H.~Hong, H.~Jiang, Z.~Zhang, K.-K. Wong, Y.~Xu, and W.~Zhang, ``Learned blockwise port activation for real-time beamforming in fluid antenna arrays,'' arXiv:2607.25365, 2026.

\bibitem{Dinis2026SpatioTemporal}
D.~Dinis and R.~Wichman, ``Spatio-temporal port selection in fluid antennas under switching delays,'' \emph{IEEE Communications Letters}, vol.~30, pp. 992--996, 2026.

\bibitem{Zhu2026HRLLC}
X.~Zhu, K.-K. Wong, H.~Xu, C.~Rao, and H.~Shin, ``Fluid antenna systems enabling 6G {HRLLC} with port switching delay,'' arXiv:2605.06275v2, 2026.

\end{thebibliography}
\end{document}